 \documentclass[final,3p,times,twocolumn]{elsarticle}

\usepackage{amssymb}

\usepackage[breaklinks=true,colorlinks=true,linkcolor=blue,urlcolor=blue,citecolor=blue]{hyperref}

\usepackage{amsfonts}
\usepackage{slashed}
\usepackage{graphicx}
\usepackage[usenames,dvipsnames]{color}
\usepackage{float}

\newcommand{\be}{\begin{equation}}
\newcommand{\ee}{\end{equation}}
\newcommand{\beq}{\begin{eqnarray}}
\newcommand{\eeq}{\end{eqnarray}}

\journal{Nuclear Physics B}

\begin{document}

\begin{frontmatter}



\title{Spontaneous mass generation and  the small dimensions \\ of the Standard Model gauge groups $U(1)$, $SU(2)$ and $SU(3)$}


\author{Guillermo Garc\'{\i}a Fern\'andez, Jes\'us Guerrero Rojas, \\
and Felipe J. Llanes-Estrada}

\address{Depto. F\'{\i}sica Te\'orica I, Universidad Complutense de Madrid, Parque de las Ciencias 1, 28040 Madrid, Spain.}

\begin{abstract}
The gauge symmetry of the Standard Model is 
$SU(3)_{c} \times SU(2)_{L} \times U(1)_{Y}$ for unknown reasons. One aspect that can be addressed is the low dimensionality of all its subgroups. Why not much larger groups like $SU(7)$, or for that matter, $SP(38)$ or E7?\\
We observe that fermions charged under large groups acquire much bigger dynamical masses, all things being equal at a high e.g. GUT scale, than ordinary quarks. Should such multicharged fermions exist, they are too heavy to be observed today and have either decayed early on (if they couple to the rest of the Standard Model) or become reliquial dark matter (if they don't).\\ 
The result follows from strong antiscreening of the running coupling for those larger groups 
(with an appropriately small number of flavors)
together with scaling properties of the Dyson-Schwinger equation for the fermion mass.\\
\end{abstract}

\begin{keyword}
Standard Model group\sep Running fermion mass\sep Grand Unification scale
\PACS 14.65.Jk \sep 11.10.Hi \sep 12.10.Kt
\MSC[2010]  81T13 \sep 81T80 \sep 22E70 

\end{keyword}

\end{frontmatter}


\section{Introduction}

The Lagrangian density of the Standard Model of particle physics features the gauge symmetry
\begin{equation}
SU(3)_{c} \times SU(2)_{L} \times U(1)_{Y}\ .
\end{equation}
At the hadronic scale, Quantum Chromodynamics (QCD, $SU(3)_c$) has evolved to a strongly coupled theory with spontaneous mass generation (and correspondingly, Chiral Symmetry Breaking) whereas the two smaller groups entail theories that remain perturbatively tractable, with small coupling.

At high energies, the non-Abelian theories become asymptotically free and all three couplings approximately merge at a large Grand Unification Theory (GUT) scale towards which also other phenomena in particle physics point.

Why these groups are symmetries of particle physics at collider energies is not obvious. One feature that calls our attention at first is the 1-2-3 succession of small numbers. Classical Lie groups can have arbitrary dimensionality. Why the first three integers?
It is fashionable to resort to anthropic reasoning, perhaps within a landscape of theories (``this symmetry group is compatible with life''), but 
there could also be more satisfactory explanations.

In this article we adopt the view that arbitrarily larger symmetries could be manifest at very high energy scales, but that fermions charged thereunder would become so massive as to be out of the reach of particle colliders.

We show that if the coupling constants $\alpha_s$ and the $O$(MeV) fermion masses are about equal for all the groups at the GUT scale $10^{15}$ GeV, and compatible with light quarks charged under $SU(3)_c$ acquiring a constituent mass of about 300 MeV (so they are phenomenologically viable in hadron physics), then fermions charged under larger groups are above the 10 TeV scale and not yet detectable. 

That is to say, fermions charged under groups of larger dimension than the Standard Model might exist, but if the coupling of those groups was similar to those of the SM at some GUT scale, those fermions are not detectable with present instrumentation.

We will show that the dynamical mass of those fermions grows exponentially with the group's fundamental dimension (for relatively small $N_c$), i.e.
\begin{equation} \label{mainresult}
M(0)_{N_c} \propto e^{N_c} \times \theta(N_f^{\rm critical} -N_f)
\end{equation}
and then increases more slowly for larger $N_c$, saturating towards the GUT scale (where all are equally light by construction).
The Heavyside step function in flavor limits the validity of the result to fermions whose flavor degeneracy is smaller than a certain critical value at which the vacuum polarization becomes insufficiently antiscreening (and beyond which dynamical chiral symmetry breaking ceases). This is further discussed below in subsection~\ref{subsec:flavor}. 

To establish the result shown in Eq.~(\ref{mainresult}), we will find rescaled solutions of the mass Dyson-Schwinger equation that allow us to avoid difficult numerical integration over large intervals of momentum. We will use these solutions in conjunction with a perturbative analysis of the highest energy scales, where $\alpha_s$ is small. The key of the analysis is to note that the scale at which the coupling constant times the relevant color factor becomes sizeable, so that the DSE needs to be employed (which for concreteness we will take as $(C_F \alpha_s) = 0.4$)
is larger for larger groups due to the increased antiscreening in Yang-Mills theories, so that the fermion mass runs for larger intervals and thus becomes much larger at $p=0$.

In section~\ref{sec:DSE} we introduce and simplify, following standard theory, the DSE for the fermion propagator. There, in subsection~\ref{subsec:cutoff}, we will already change the group under which the fermions are charged and observe, numerically and at fixed cutoff, that the solutions for larger groups seem to be simple rescalings of the known $SU(3)$ solution. In subsection~\ref{subsec:MOM} we will change to the MOM scheme to avoid the inconvenients of cutoff solutions. Section~\ref{sec:per} takes us to the highest energies where the use of perturbation theory is appropriate, and we will briefly recall antiscreening and perturbative mass running in non-Abelian Yang-Mills theories.

The crux of the article is then section~\ref{sec:both}, where the scaling properties of the rainbow-ladder DSE are combined with the perturbative analysis to yield our main result, shown in figure~\ref{fig:mofNc}: that the fermion mass becomes very large for larger groups, and that it scales for moderate $N_c$ as in Eq.~(\ref{mainresult}).
Further discussion spans section~\ref{sec:outlook}. The appendix is reserved for mathematical detail (computation of the group color factor $C_F$ are reported there).

\section{Some properties of spontaneous mass generation}\label{sec:DSE}
The mass function plays a central role in gauge theories coupled to fermions and their uses for phenomenology.
A brief summary discussing several subtleties and identities is given in~\cite{Bashir:2012fs}.

We want to adopt the simplest possible Lorentz-invariant model that exposes the physics. The Nambu-Jona-Lasinio model is a practical option to demonstrate spontaneous mass generation, but its contact-interaction structure cannot be used at high energies, where the coupling is not transparently related to the running coupling of the underlying non-Abelian theory.

Next in difficulty is the rainbow approximation to the Dyson-Schwinger equation~\cite{Roberts:2012sv} of the fermion propagator in the gauge theory, so we settle to it~\cite{Blaschke:2007ce}. While a very basic approximation, the simplicity of the scenario we propose does not require more sophisticated many-body methods. Rainbow-ladder approximation is still widely used for exploratory studies of beyond the standard model physics~\cite{Matsuzaki:2015sya}.

\subsection{Dyson-Schwinger equation for a fermion propagator}\label{subsec:DSE}

The free propagator of a fermion with current mass $m_c$ is denoted as
\begin{equation}
S_{0}(p^{2})= \frac{i}{ \slashed p - m_{c}}\ .
\end{equation}
The full propagator is usually parametrized as
\begin{equation}
S(p^{2})= \frac{i}{A(p^{2}) \slashed p - B(p^{2})}.
\end{equation}
but to expose spontaneous mass generation it is sufficient to consider a simplified ansatz with $A(p^{2})=1$ and running mass $B(p^2)=M(p^2)\equiv M_{p}$.

\begin{figure}[t]
\centering
\includegraphics*[width=8cm]{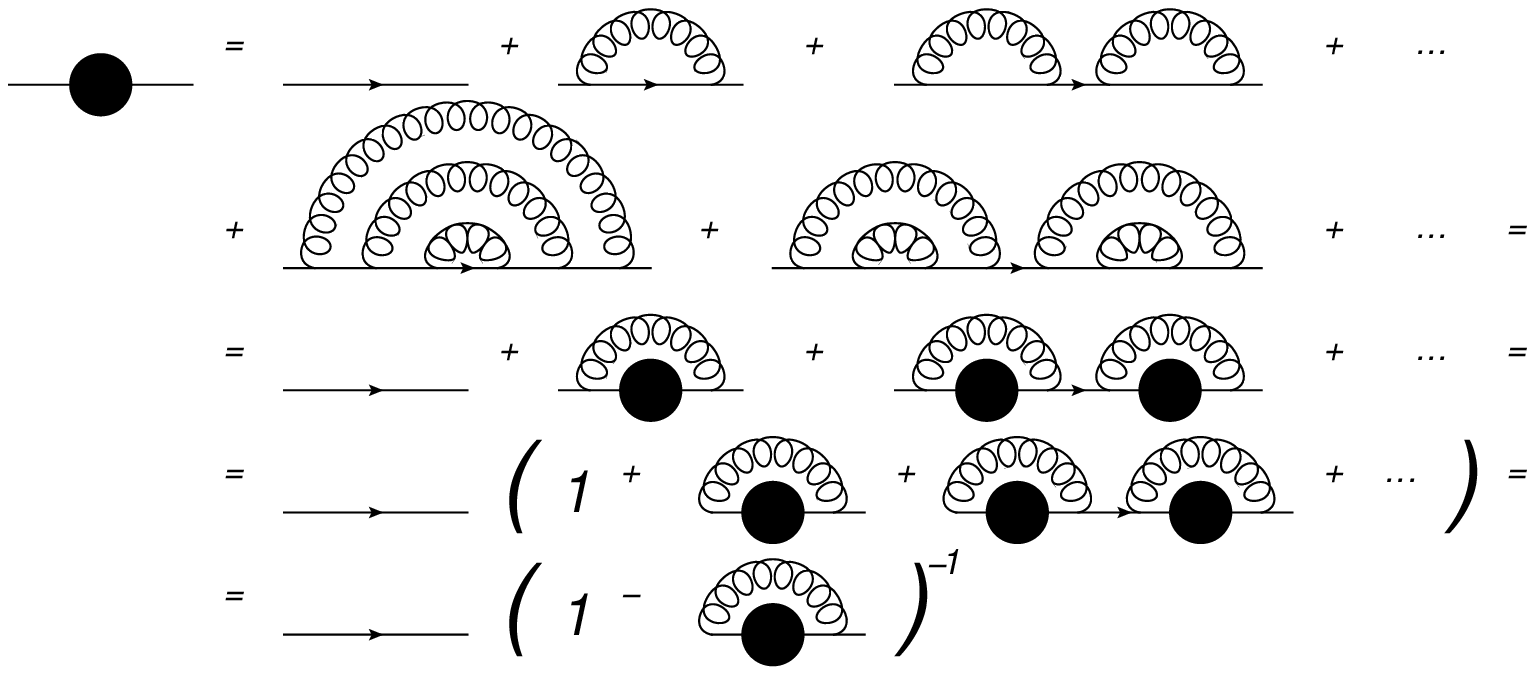}\vspace{0.5cm}
\includegraphics*[width=5.5cm]{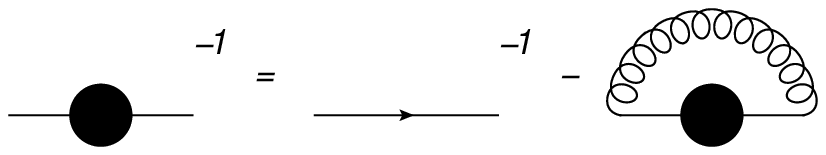}
\caption{Resummation of the rainbow diagrams (with perturbative gauge boson propagator and fermion-boson vertex) leading to the DSE for the fermion propagator in rainbow approximation. \label{fig:rainbow}}
\end{figure}

The Dyson-Schwinger equation (DSE) for this full propagator,
\begin{equation}
S^{-1}(p^{2})=S^{-1}_{0}(p^{2})-\Sigma(p^{2})
\end{equation}
may be written down as an identity in the field theory, but can pedagogically be deduced as a resummation of perturbation theory. The rainbow resummation avoids all diagrams with vertex corrections, counting only those of the type depicted in figure~\ref{fig:rainbow}. 

After standard manipulations~\footnote{Tracing over Dirac matrices, performing a Wick rotation to Euclidean space $q^{0} \rightarrow iq^{0}$, $p^{0} \rightarrow ip^{0}$, $\int{d^{4}q} \rightarrow i\int{d^{4}q_{E}}$, and employing 4D spherical coordinates.}, 
the DSE takes the well-known form
\begin{equation}
M_{p}=m_{c}+\frac{C_F}{\pi^{3}}\int _{0}^{ \infty}{ q^{3}dq\frac{M_{q}}{\left| q \right|^{2}+M^{2}_{q}}g^{2}D^{0}_{p-q} }. \label{DSE}
\end{equation}
where $C_F$ is the color factor (or Casimir of the group's fundamental representation) which is the object that we will vary in this investigation.
Also seen are \textit{g}, the fermion (non-Abelian) charge; and 
the Feynman-gauge gauge-boson, or for short even beyond QCD, ``gluon'' propagator
\begin{equation}\label{bareprop}
-iD((p-q)^{2})\eta_{\mu \nu} =  \frac{-i \eta_{\mu \nu}}{(p-q)^{2}}
\end{equation}
averaged over 4-dimensional polar angle, 
\begin{equation}
\int _{-1}^{1}{ \sqrt{1-x^{2}}}D_{p-q}dx \equiv D^{0}_{p-q}\ .
\end{equation}
This ($N_c$-independent) gauge boson propagator is taken to be perturbative, though if need be, this can be corrected in future work to achieve better precision (see~\cite{Huber:2013yqa} for a very brief outline of the current estimates in non-Abelian gauge theory, and~\cite{Aguilar:2015nqa} for more extended discussion). The use of the same propagator for all $N_c$ is supported by independent studies~\cite{Maas:2010qw}.

To solve the DSE we discretize the variables $p$, $q$ and the function $M$, so the $q$-radial and $x$-angular integrals become discrete sums (needing regularization as they are divergent at large $q$), and linearize  $M= M_{0}+m$ where $M_0(p^2)$ is a guess and $m(p^2)$ the unknown correction returning the correct solution $M(p^2)$. 
Expanding Eq.~(\ref{DSE}) to first order in $m$ provides a linear system for $m(p^2)$ solved with a linear algebra package. The improved $M(p^2)$ is used as a new guess $M_0(p^2)$ and the procedure iterated until $m\simeq 0$.

\subsection{Mass generation at the hadron scale (with cutoff regularization)}
\label{subsec:cutoff}

To show the reaction of the DSE Eq.~(\ref{DSE}) to changing the group, we first study the hadronic scale cutting off the $q$ integral at $\Lambda=10$ GeV.
We take the (cutoff-dependent) current mass $m_c=m(\Lambda^2)=0$ for the free fermion to vanish, and solve for $M(p^2)$ at smaller scales, so the entire mass function is here dynamically generated breaking the global chiral symmetry. 

\begin{figure}[t]
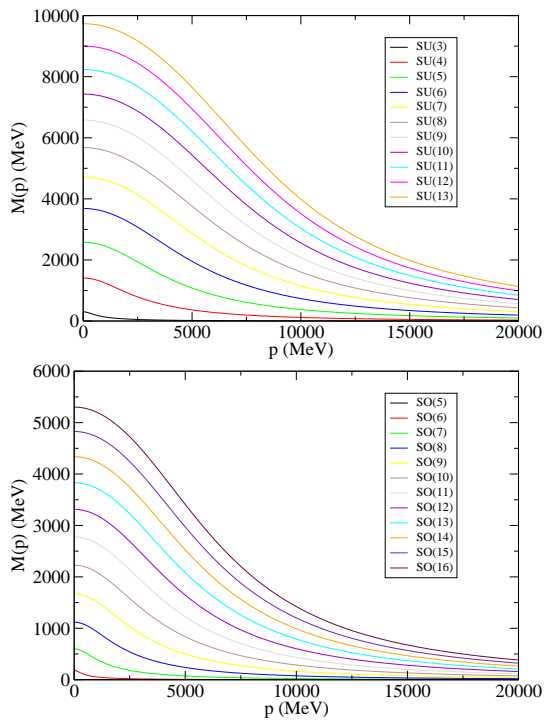

\centering
\includegraphics*[width=7cm]{FIGS.DIR/SUcutoff.eps}
\includegraphics*[width=7cm]{FIGS.DIR/SOcutoff.eps}
\caption{Mass function for the special unitary SU($N_{c}$) (top) and orthogonal SO($N_c$) (bottom) with momentum integral regularized at $ \Lambda=10$GeV.
In this fixed momentum interval the constituent mass grows nearly linearly with the group dimension (of the fundamental representation). \label{fig:cutoff1}}
\end{figure}

\begin{figure}[t]
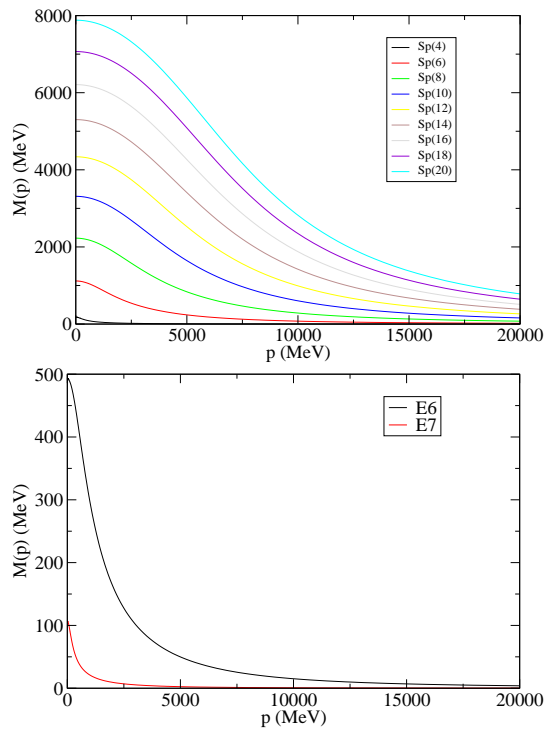

\centering
\includegraphics*[width=7cm]{FIGS.DIR/Spcutoff.eps}
\includegraphics*[width=7cm]{FIGS.DIR/SGcutoff.eps}
\caption{Mass function for the symplectic groups Sp($N_{c}$) (top) and 
a couple of special groups as indicated (bottom) with momentum integral regularized at $ \Lambda=10$GeV. Again, in this fixed momentum interval the constituent mass grows nearly linearly with the group fundamental dimension as in fig.~\ref{fig:cutoff1} \label{fig:cutoff2}.}
\end{figure}

To be specific, in the calculations shown in figures~\ref{fig:cutoff1} and~\ref{fig:cutoff2}, the coupling $g$ is taken to be the same for all groups and fixed by demanding that the quark mass for $SU(3)$ be 300 MeV, as corresponds to the observed QCD quarks. This results in a value $g\simeq 15.1$ with the cutoff fixed at 10 GeV.

 This value of $g$ amounts to $\alpha_{s}\simeq 18$, much larger than one naively expects in QCD. This is due to several reasons, among them having set $m_{c}=0 $, which suppresses $M(0)$ a moderate amount; having fixed $\Lambda=10$ GeV, which restricts  the range of running mass a bit; and saliently, the use of a bare $qqg$ vertex. Since chiral symmetry breaking has to be simultaneous in all Green's functions~\cite{Alkofer:2008tt}
and there is feedback between them, our use of a bare vertex underestimates the extent of chiral symmetry breaking, requiring a larger $\alpha_s$ for equal $M(0)$. 

We can think of this larger $g$ as simply the product $gV$ with $V$ a vertex strength factor. For $SU(3)$ this factor is $7.7$, and for other groups it scales as $V(N_{c})=V_{SU(3)}\frac{N_{c}}{3}$, which is the leading $N_c$ behavior of the vertex one-loop corrections (specifically, the non-Abelian correction). There is a vast literature on vertex corrections that dates back decades, see e.g.~\cite{Ball:1980ax} or~\cite{Atkinson:1992tx}, so we abstain from further investigation here as this would carry us too much off topic. But it is clear that the rainbow-ladder approximation is just a first approximation to the physics.

The results for a couple of special groups and for all the classical Lie groups (SU($N_c$), SO($N_c$) and Sp($N_c$), with $N_c$ being even for the later)  clearly show mass functions that seem to be  rescalings of one another upon changing the group dimension~\footnote{we will elaborate on this property later in section~\ref{sec:both}.}, with mass generation almost directly proportional to the fundamental dimension of the group.

\begin{figure}[hbtp]
\centering
\includegraphics*[width=8cm]{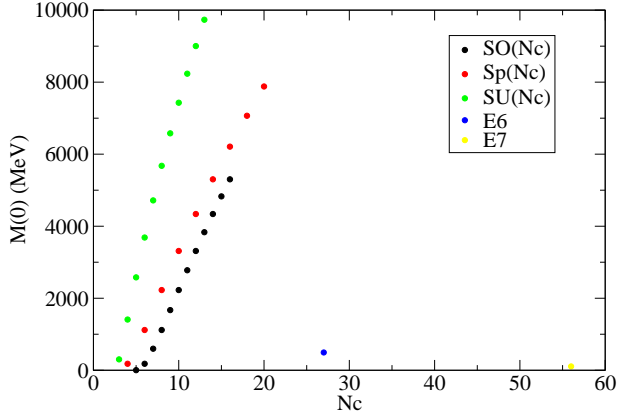}  
\caption{Dependence of the constituent mass $M(0)$ with the color number $N_{c}$ under a cutoff regularization with $\Lambda=10$GeV. For a given classical group family, the dependence is rather linear. \label{fig:mcutoff}}
\end{figure}

Let us now concentrate on the ``constituent'' mass $M(0)$ seen at lowest energies, while varying the color number $N_c$ and the group families. For this we extract the first point of each $M(p^2)$ function and plot the outcome in figure~\ref{fig:mcutoff}.

From the figure, it stands out that for $U(1)$ and $SU(2)$ there is no chiral symmetry breaking, i.e. $M(0)=0$, for the same coupling intensity that generates the 300 MeV quark mass in $SU(3)$.
Past dedicated $SU(2)$ (and also $G2$) lattice studies\cite{Maas:2010qw,Maas:2007af} found that the general structure of the Green's functions is similar to the $SU(3)$ case, for commensurate but larger coupling (presumably to make up for the reduced color factors) {\it{at a low, hadronic scale}}. Our setup, and thus our result, differs in that the couplings are equal {\it{at a high-energy scale}} so the coupling for smaller groups is much smaller at the lower scale.

A related, dedicated study~\cite{Hopfer:2014zna} shows how lowering the antiscreening of QCD eliminates dynamical mass generation.

Beyond $U(1)$ and $SU(2)$, we find no mass generation for $G_2$ ($N_c=7$) and $F_4$ ($N_c=26$), both with a relatively small color factor  $C_F$=1 in spite of their large dimension; and also for $SO(N_{c})$ with  $N_{c}$=1 to 5 and for $Sp(2)$.
For all these groups, an explicit fermion mass $m$ just yields an $M(p^2)$ that slightly separates from the perturbative value without really yielding symmetry breaking. 

For the rest of the classical groups, where symmetry breaking is apparent, the dependence of $M(0)$ on the defining dimension $N_c$ is seen to be rather linear. This, as we will see, happens because we have integrated over the same momentum interval $(0,\Lambda)$. 

\begin{figure}[hbtp]
\centering
\includegraphics*[width=8cm]{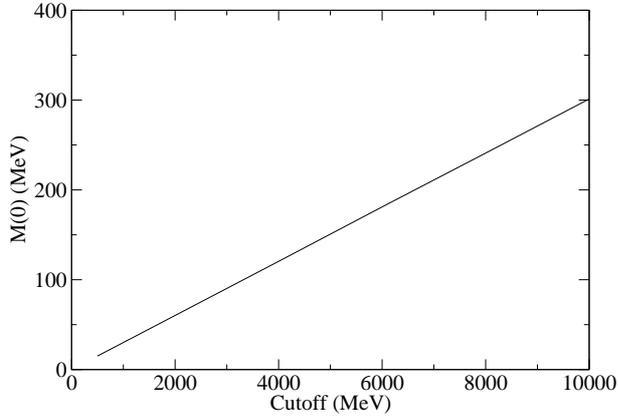}  
\caption{Dependence of the $SU(3)$ constituent mass $M(0)$ with the cutoff.  
This very nicely linear relation would get modified in a more sophisticated truncation of the gauge theory where the gluodynamics generates an additional scale (a gluon mass-like parameter that cuts the propagator in the infrared, another topic on which there is a large literature). We stay with a strictly massless gauge boson propagator as in Eq.~(\ref{bareprop}) through the entire article.
\label{fig:mcutoff2}}
\end{figure}

From the linear dimension of the leading divergence in the DSE one can also deduce that $M(0)\propto\Lambda$, which can anyway be checked numerically as shown in figure~\ref{fig:mcutoff2}.

After this warmup, we have shown that mass generation at the hadron scale is insufficient to expel fermions charged under large groups from the spectrum. This is no longer true when considering
high-energy physics, where running over large momentum swaths is involved.
 But before proceeding, we note that cutoff regularization is inadequate (now that the highest scale will be pushed to $10^{15}$ GeV), so we first introduce an appropriate renormalization scheme in the next subsection.

\subsection{One technical improvement: momentum subtraction scheme}
\label{subsec:MOM}

There are many reasons to improve on simple cutoff regularization, among them preserving Lorentz invariance and exposing renormalizability. To characterize the quantized theory we need a renormalization scale $\mu$ at which the couplings $ \alpha_{s}\equiv g^{2}/4\pi$ are to be chosen. To achieve this, we will adapt a variation of the Momentum Subtraction Scheme or MOM often used in this subfield of Dyson-Schwinger equations.  Since we will later, in our perturbative analysis, employ only 1-loop running of masses and coupling constants, we can take the renormalization group coefficients $\beta$ and $\gamma$ to be the same as in the more usual Modified Minimal Subtraction Scheme ($\overline{\rm MS} $), as they are equal to one loop
(see~\cite{Almeida:2010ns}).

The first step is to introduce adequate renormalization $Z(\Lambda^2,\mu^2)$-constants that absorb any infinities or, once regulated, any dependence on the cutoff $\Lambda$,
\beq \nonumber
S^{-1}(p^{2},\mu^{2})\equiv Z_{2}S^{-1}_{0}(p^{2})-\Sigma(p^{2},\mu^{2})\ ,
\\ \nonumber
\Sigma\equiv ig^{2}C_F\int { \frac{d^{4}q}{(2\pi)^{4}} \gamma^{\mu} S(q^{2},\mu^{2}) \gamma_{\nu} D((p-q)^{2},\mu^{2}) }\ , \\
\eeq
namely $Z_2$ for the wavefunction renormalization and $Z_m$ for the bare quark mass. We do not calculate vertex corrections nor loops involving ghosts in this article since they are an unnecessary complication for the physics exposed, so we need no additional $Z$ constants beyond those of the bare quark (inverse) propagator, $S^{-1}_{0}(p^{2}) $.
Therein, the relation between the (cutoff dependent) unrenormalized mass
$ m_{c}(\Lambda^{2})$ and the renormalized mass at the renormalization scale
$m_{R}(\mu^{2})$ is~\cite{Fischer:2003rp}  
\begin{equation}
m_{c}(\Lambda^{2})=Z_{m}(\Lambda^2,\mu^2) m_{R}(\mu^{2})\ .
\end{equation}
Should we lift the restriction $A=1$, the renormalization of the wavefunction would entail $A^{-1}_{0}(p^{2},\Lambda^{2})=Z_{2}A^{-1}(p^{2},\mu^{2})$; though while we maintain it, then also $Z_2=1$ and the only needed renormalization condition is to fix the mass at $p^2=\mu^2$. The DSE  for the mass function is then formally
\begin{equation}
M(p^{2})=Z_{m}m_{R}(\mu^{2})+\Sigma_{_{M}}(p^{2},\mu^{2})\ ;
\end{equation}
evaluating it at $p^2=\mu^{2}$ and subtracting both, we obtain
\begin{equation}
M(p^{2})=M(\mu^{2})+\Sigma_{_{M}}(p^{2},\mu^{2})-\Sigma_{_{M}}(\mu^{2},\mu^{2}),
\end{equation}
in terms of finite quantities alone. Thus, the resulting MOM equation is
\beq
M(p^2)=M(\mu^2)+\frac{g^2C_F}{\pi^3}\int_0^\infty q^{3}dq
\nonumber \\ 
\frac{M(q^2)}{\left| q \right|^2 + M^2(q^2)}(D^0_{p-q}-D^0_{\mu-q})
\eeq
(with $\mu$ parallel to $p$), that is,
\beq  \label{subtractedDSE}
M(p^2)= M(\mu^2) + \frac{g^2C_F}{\pi^3}\int_0^\infty q^3dq
 \int_{-1}^1dx \sqrt{1-x^2}  \nonumber \\
\left( \frac{1}{ \left| q  - p \right|^2 }- 
       \frac{1}{ \left| q  -\mu\right|^2 } 
\right)  
       \frac{M(q^2)}{M^2(q^2)+ \left| q \right|^2 }.
\eeq
If the $q$ radial integral in this equation is cutoff at $\Lambda>>(\mu,p)$, it is easy to see that asymptotically, for $\mu$ and $p$ parallel,
\beq
\frac{\partial M(p^2)}{\partial \Lambda}\propto  \frac{M(\Lambda^2)(p-\mu)}{\Lambda^2}
\eeq
so that for large $\Lambda$ and $M$ growing slower than quadratically at large momentum, $M(p^2)$ stops depending on the cutoff, renormalization is achieved and $M(\mu^2)$ alone determines the function for values of $p$ smaller than $\mu$.

\begin{figure}[t]
\centering
\includegraphics[width=7cm,angle=-90]{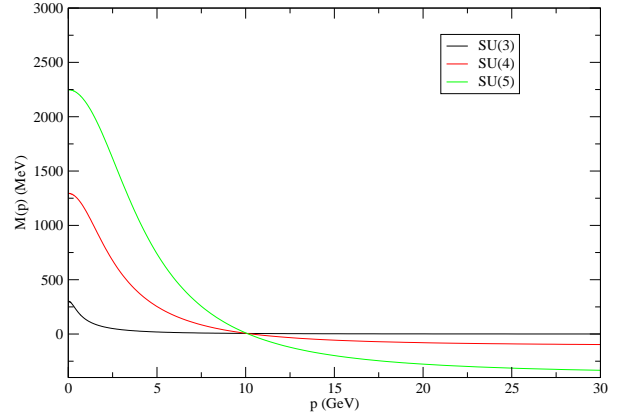} 
\caption{Mass functions for some $SU(N_c)$ groups in the MOM scheme with 
renormalization point $\mu=10$GeV, 
where we have chosen $M(\mu^{2})=M_{SU(3)}(\mu^{2})$, the latter such that
$M_{SU(3)}(0)=300$ MeV. To compare the very different growth of the three mass functions even at low scales, we have chosen them equal at a very low $p$ so that the $SU(4)$, $SU(5)$ ones eventually become negative at high energies. This is of course unphysical, and just means that $M(\mu)$ should naturally be chosen larger because chiral symmetry is already broken. We nevertheless find the plot instructive.
\label{fig:MOM} }
\end{figure}

We again fix (for all groups) $g=15.07$ at $\mu=10$ GeV so that for $SU(3)$ the constituent quark mass is $M(0)=300$ MeV once more. 
We impose the renormalization condition $M(\mu^{2})_{SU(3)}=5.7$ MeV for all groups. The tail of the mass function for the group SU(3) approaches zero asymptotically, as shown in figure~\ref{fig:MOM}. 

Also shown are mass functions for $SU(4)$ and $SU(5)$ that are seen to change sign. This is not necessarily that the computer code has found the excited, sign-changing solutions of~\cite{Bicudo:2002eu,Nefediev:2002tb,Martin:2006qd}. 
Instead, what it shows is that the self-energy at $\mu$ subtracted in
Eq.~(\ref{subtractedDSE}) is very large and overcomes the smaller self-energy computed at $p$ as well as the smaller mass chosen at 5.7 MeV. This simply reflects a renormalization point $\mu$ that is too low for the higher groups, before the perturbative behavior sets in (but we want to compare the three functions at the same point), so we are not only subtracting the ultraviolet divergence but also large finite-$p$ contributions.
 This suggests, as we will soon effect, to move the renormalization point of the larger groups to a much higher scale where the coupling is weaker. 

Ignoring that sign for now, the solutions are seen to be similar in shape to the ones obtained with the cutoff method.  
Turning now to the deep infrared, we conclude that the outcome is equivalent to that obtained in subsection~\ref{subsec:cutoff}, with $M(0)$ scaling in proportion to $N_c$ if only the hadron scale is considered, so we have achieved a very simple renormalization that allows us to proceed to higher scales.

\section{Treatment of the high-energy running mass within perturbation theory}
\label{sec:per}

We now extend our study to the Grand Unified Theory scale at $10^{15}$GeV. 
Several physics coincidences point out to some dynamics taking place at that scale, for example the see-saw Majorana mass scale in neutrino physics, and most important for this work, the approximate coincidence of the coupling constants of the Standard Model gauge interactions at that scale (see~\cite{Agashe:2014kda} for an introductory review).

\subsection{Running coupling and mass}
In that energy regime, the running of the mass and coupling constants can be followed in perturbation theory, as long as $\alpha_s$ remains small.
Up to one loop, we will need the $ \beta_1$ and $\gamma_1$ coefficients of 
the $\beta$-function and of the anomalous mass dimension, respectively
\beq \label{renormgroup1}
\beta(a_{s}) \equiv -\mu\frac{da_{s}}{d\mu}
\nonumber\\
= \beta_{1}a^{2}_{s}+\beta_{2}a^{3}_{s}+...
\eeq
with $a(s)= \frac{\alpha_s}{\pi}$, and
\beq \label{renormgroup2}
\gamma(a_{s}) \equiv -\frac{\mu}{m}\frac{dm}{d\mu}
\nonumber \\
= \gamma_{1}a_{s}+\gamma_{2}a^{2}_{s}+...
\eeq
We will, for simplicity of the argument, consider that there is only one fermion flavor charged under each of the color groups, so that we may set $N_f=1$. Following~\cite{Jaminlectures,Muta}, we have
\beq
\beta_{1} &=&\frac{1}{6}(11N_{c}-2N_{f})\ , \label{beta1}\\
\gamma_{1}&=&\frac{3}{2}C_F\ ,
\eeq
and the solutions to eqs.~(\ref{renormgroup1}) and~(\ref{renormgroup2}) is obtained after integrating once,
\begin{equation}
\int _{a_{s}(\mu_{1})}^{a_{s}(\mu_{2})}{\frac{da_{s}}{\beta(a_{s})}}= \ln { \frac{\mu_{1}}{\mu_{2}} }, 
\end{equation}
\begin{equation}
\int _{a_{s}(\mu_{1})}^{a_{s}(\mu_{2})}{{da_{s}}\frac{\gamma(a_{s})}{\beta(a_{s})}}=\ln { \frac{m(\mu_{2})}{m(\mu_{1})} }\ ,
\end{equation}
from which follow the well known forms
\begin{equation} \label{runningalpha}
\alpha_{s}(\mu_{2})=\alpha_{s}(\mu_{1}) \frac{1}{1+\frac{\alpha_{s}(\mu_{1})}{\pi}\beta_{1}\ln {\frac{\mu_{2}}{\mu_{1}}}},
\end{equation}
and
\begin{equation} \label{runningmass}
m_{s}(\mu_{2})=m_{s}(\mu_{1}) \left( \frac{1}{1+\frac{\alpha_{s}(\mu_{1})}{\pi}\beta_{1}\ln {\frac{\mu_{2}}{\mu_{1}}}} \right)^{\frac{\gamma_{1}}{\beta_{1}}}\ .  
\end{equation}

In what concerns our study, it is worth remarking that groups of equal dimension in different families have differently running masses (for equal and low flavor number, in our estimates $N_f=1$). This is in spite of {\emph{the running of $\alpha_s$}} depending on the chosen group \emph{only} through its defining dimension $N_c$ (of course, equal to the adjoint Casimir $C_A$). The reason is that the actual color factor that appears exponentiating the fermion mass in Eq.~(\ref{runningmass}) is the Casimir $C_F$ of the fundamental representation, which is different for two groups belonging to different families even if they have the same dimension (in short, equal $N_c$ does not imply equal $C_F[N_c]$).

\begin{figure}
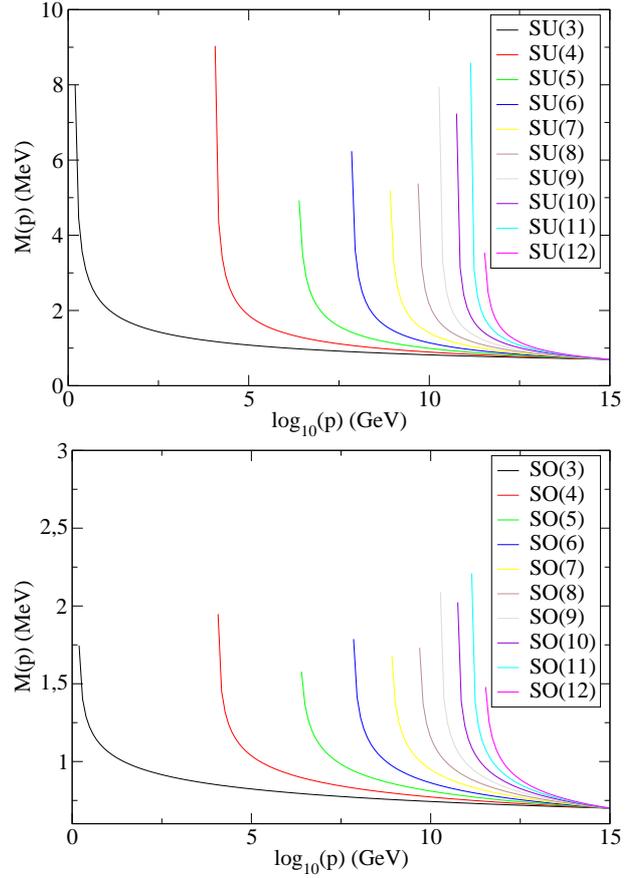

\centering
\includegraphics*[width=8cm]{FIGS.DIR/SUmasahe.eps}
\includegraphics*[width=8cm]{FIGS.DIR/SOmasahe.eps}
\caption{Running mass $M(p^2)$ as a function of $N_c$ from 
perturbation theory running at one loop from the GUT scale. Here we depict the classical unitary and orthogonal group families. 
\label{fig:masahe1}}
\end{figure}

\begin{figure}
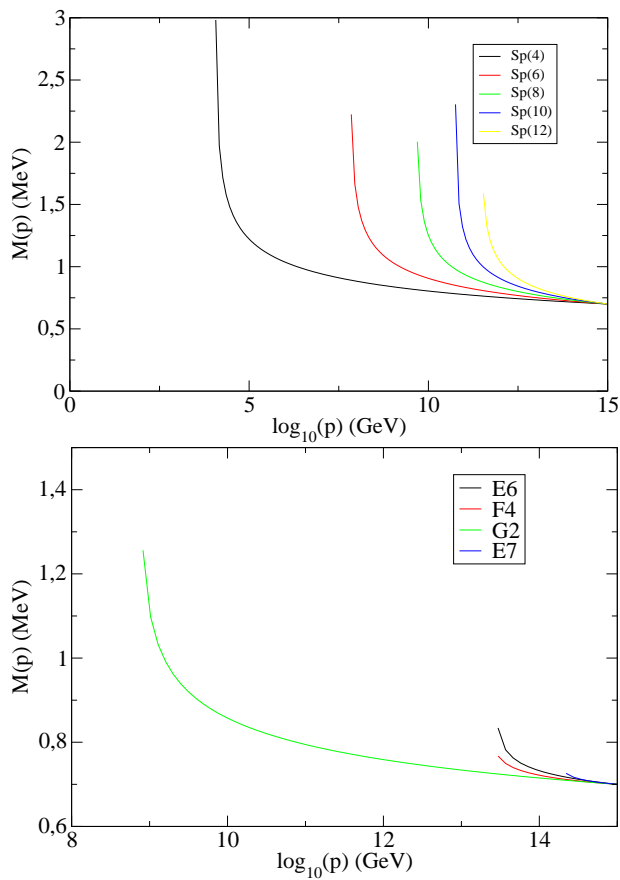

\centering
\includegraphics*[width=8cm]{FIGS.DIR/Spmasahe.eps}
\includegraphics*[width=8cm]{FIGS.DIR/SGmasahe.eps}
\caption{Running mass $M(p^2)$ as a function of $N_c$ from 
perturbation theory running at one loop from the GUT scale. Here we depict the classical symplectic group family and some exceptional groups.
 \label{fig:masahe2}}
\end{figure}

These running masses are depicted in figures~\ref{fig:masahe1} and~\ref{fig:masahe2}, that already hint at much heavy fermions even in perturbation theory.

Returning to the running coupling, $\beta_{1} $ is positive for non-Abelian Yang-Mills theories ($N_c\neq 1$), so $ \alpha_{s}(\mu_{2}) $ decreases logarithmically for $ \mu_{2}>\mu_{1} $, and asymptotic freedom is manifest. 
Running in the opposite direction towards lower energies, the intensity of interaction increases until a Landau pole  $ \Lambda $ is hit (not to be confused with the earlier cutoff), where the denominator of Eq.~(\ref{runningalpha}) vanishes,
\begin{equation}
\Lambda=\mu_{1}{e}^{-\frac{1}{\beta_{1}a_{s}(\mu_{1})}}\ .
\end{equation}
Much earlier than that pole, these analytical formulae cease to be applicable
and must be substituted by resummation, e.g. by DSEs.
The Landau pole is of course a notorious feature of perturbation theory, that 
is avoided in other approaches. In Analytical Perturbation Theory, for example, $\alpha_s$ saturates at low energies~\cite{Shirkov:2006gv};
Dyson-Schwinger equations studying the gluon-ghost sector of Landau-gauge QCD concur~\cite{Alkofer:2004it}; and generally one does expect a flattening of $\alpha_s$ at low scales, yielding a conformal window~\cite{Brodsky:2011ig}.

Therefore we need to match the high-energy treatment, that can be handled in perturbation theory as just explained, with the earlier DSE treatment at some scale  $ m(\mu^{2}) $, which is the object of the next section.

\subsection{Effect of the number of flavors} \label{subsec:flavor}

The reader will have noticed that Eq.~(\ref{beta1}) depends on the number of flavors, which we have taken as $N_f=1$ for the numerical examples (in lattice language, this is the ``quenched approximation''). However, as it is well known, 
if there is a sufficiently large fermion degeneracy, which in one-loop perturbation theory as encoded by that equation is $N_f = \frac{11 N_c}{2}$, the vacuum polarization becomes screening instead of antiscreening (the sign of $\beta_1$ changes). 

The number of flavors necessary for this screening in $SU(3)$ is 17, and for $SU(4)$ it is 22, and larger yet for higher groups, which seems a rather large degeneracy. However, for smaller $N_f$ one may have an antiscreening, yet too weak, interaction that will fail to trigger dynamical chiral symmetry breaking and thus a nonperturbative fermion mass. 

Estimates of the critical flavor number beyond which chiral symmetry breaking ceases have been provided in the literature. Closest in spirit to our work are those from the DSEs~\cite{Bashir:2013zha} as well as the Renormalization Group Equations~\cite{Braun:2010qs}. 
The DSE estimate in~\cite{Bashir:2013zha} is, for $SU(3)$, $N_f^{\rm critical} = 8\pm 1$. The second work quotes numerical estimates that are compatible within the error, $N_f^{\rm critical} =11\pm 2$.

Because of Eq.~(\ref{beta1}), it is plausible that $N_f^{\rm critical}\propto N_c$, so that the number of flavors necessary to overturn chiral symmetry breaking keeps growing (so that, for example, for $N_c=4$ we would have $N_f^{\rm critical} = 11 \pm 2$ or $15\pm 3$ respectively).

The existence of this critical number of flavors justifies the $\theta(N_f^{\rm critical} - N_f)$
factor in Eq.~(\ref{mainresult}): above that number, $M(0)$ becomes of order the current mass $m_c$ and depends only radiatively on $N_c$. $N_f$ acts as the parameter of a quantum phase transition and our results apply only to the broken symmetry phase.

As a digression, for $N_f$ close but above $N_f^{\rm critical}$, our rainbow-ladder approximation in section~\ref{sec:DSE} yields Miransky scaling (see {\it e.g.} \cite{Miransky:1996pd}), by which $M(0)\propto \Lambda \exp((\rm const.)/(\sqrt{N_f-N_f^{\rm critical}}))$. Beyond rainbow-ladder, this exponential becomes modified to a power-law; the window above $N_f^{\rm critical}$ during which these critical behaviors are active is however very small~\cite{Braun:2010qs} about a few percent of $N_f^{\rm critical}$ (see fig. 5 of that work),  so that for $N_f=N_f^{\rm critical}-1$ we can safely consider ourselves in the broken phase. 

In conclusion of this subsection, though most of the considerations in this article are for a flavor-nondegenerate fermion charged under the various Lie groups, they can actually be extended to $N_f$ of modest size below $N_f^{\rm critical}$.

\section{Mass running from both high and low energies} \label{sec:both}

\begin{figure}
\centerline{\includegraphics*[width=8cm]{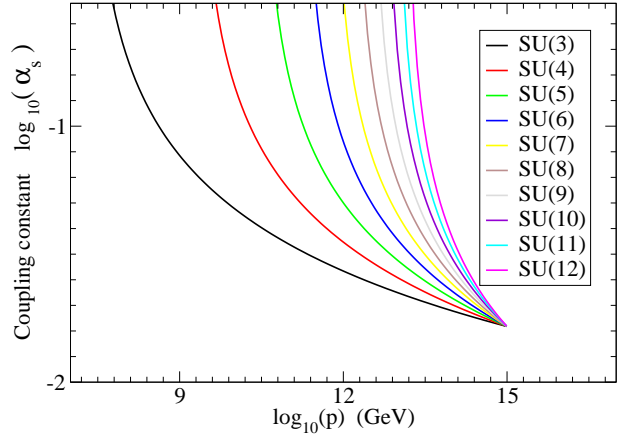}}
\caption{One-loop running coupling for the $SU(N_c)$ ($N_c\geq 3$, $N_f=1$) family of special unitary groups. For other families the running is identical since Eq.~(\ref{runningalpha}) depends, through $\beta_1$, only on the group fundamental dimension. All couplings are chosen to be identical at the GUT scale $10^{15}$ GeV.
\label{fig:runningalpha}}
\end{figure}

In this section we seek to combine the perturbative running at large scales with the DSEs at lower momenta, to obtain a picture which, even if crude, is global and allows a general statement to be produced.
We start the perturbative renormalization group running at $ \mu_{GUT}=10^{15}$GeV, where we fix
\begin{equation}
\alpha_{s}(\mu_{GUT}) = 0.017\ ,\ \ \ 
m(\mu_{GUT}) = 1MeV\ .
\end{equation}
This fermion mass is chosen to broadly reproduce the value
of the $SU(3)$-colored quark mass, that under isospin average, is taken~\cite{Agashe:2014kda} to be about
\begin{equation}
\bar { m} (2GeV) =\frac{m_{u}(2GeV)+m_{d}(2GeV)}{2} \simeq 3.5MeV\ .
\end{equation}
As for the coupling constant, the one corresponding to $SU(3)$ is precisely known at the $Z$-boson scale, $ \mu=M_{z} \simeq  100$GeV (91.2GeV), where
$\alpha_{s}(M_{z}) \simeq 0.12$.
Running to one loop and with only one fermion flavor charged under each group (shown in figure~\ref{fig:runningalpha})  requires  
an $\alpha_{_{s}}$ at the GUT scale  that is somewhat smaller than the usually quoted value $\alpha_{s}(GUT) \simeq 0.025$. But all together we seem to differ by a moderately small factor which does not affect our main argument.

We use the perturbative formulation encoded in Eq.~(\ref{runningalpha}) from $\mu_1=\mu_{GUT}$ down to $\sigma\equiv \mu_2$  where $\sigma$ represents the point where perturbation theory breaks and non-perturbative methods are required.
For $SU(3)$, this point is characterized  by $\alpha_s=0.3$, where we decide that perturbation theory must break down quickly. 
The actual combination appearing in the DSE is $g^2 C_F\propto \alpha_s C_F$. Therefore, $\alpha_s=0.3$ for $SU(3)$ is equivalent to $C_F\alpha_s= \frac{4}{3}\times 0.3=0.4$.
From that point on, we freeze $\alpha_s$ to a constant value and employ Dyson-Schwinger methods to treat the fermion mass.

\begin{figure}[t]
\centering
\includegraphics*[width=8cm]{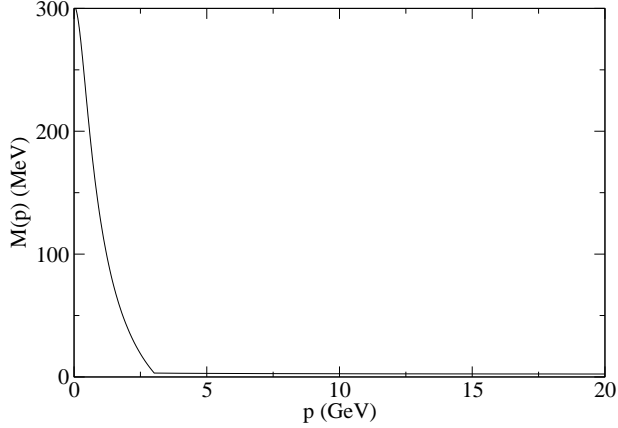}
\caption{Mass function for the SU(3) group obtained matching the numerical solution of the DSE to the perturbative one for $\alpha_s(\sigma)=0.3$.\label{fig:SU3fullrun}}
\end{figure}

In fig.~\ref{fig:SU3fullrun} we represent our complete approximation for the quark mass function in $SU(3)$. We match the perturbative and DSE solutions continuously (obtaining a smoother matching is possible by employing resummed perturbation theory on the high energy side~\cite{Alkofer:2003jj}).

If we now increase the dimension of the group $G$, the matching point $\sigma$ where $C_F \alpha_s(\sigma) =0.4$ and a non-perturbative treatment starts to be required moves much to the right of the plot to higher scales,
\begin{equation} \label{saturationpoint}
\sigma=\mu_{GUT} \times {e}^{\frac{\pi}{\beta_{1}} \left( \frac{1}{\alpha_{s}(\sigma)}-\frac{1}{\alpha_{s}(\mu_{GUT})}\right) } \ .
\end{equation}
The exponent being negative and proportional to $N_c^{-1}$, increasing $N_c$ moderately provokes an exponential increase in the scale. When $N_c$ becomes large, $\sigma\to \mu_{GUT}$ saturates and basically all further groups require non-perturbative treatment from early on.

Integration to such large scales with an appropriate grid is time consuming; it can be avoided by noticing, for example after a glance at figures~\ref{fig:cutoff1} and~\ref{fig:cutoff2}, that given a solution to the DSE's, one can easily find rescaled solutions. In those figures the color factor induced the rescaling, but now the rescaling will rather be forced by $\sigma$, the point where we start numerical integration towards lower values of $p$.

We will obtain solutions for groups of large dimension from that of $N_c=3$ shown in figure~\ref{fig:SU3fullrun}. To show that this is possible analytically, perform a scale transformation
\beq \label{transfescala}
p^{2} \rightarrow \lambda^{2}p^{2} \nonumber \\
\sigma^{2} \rightarrow \lambda^{2}\sigma^{2},
\eeq
on the DSE, where $\lambda$ is a contraction factor that will map the mass function of an arbitrary group to that of $SU(3)$.
We can always change the dummy integration variable $q^{2} \rightarrow \lambda^{2}q^{2}$, and the integration measure picks up a Jacobian $d^{4}q \rightarrow \lambda^{4}d^{4}q$.
 
With this rescaling, the DSE in Eq.~(\ref{subtractedDSE}) becomes 
\beq
\tilde M(\lambda^2p^2)= \tilde M(\lambda^2\sigma^2)
+ \frac{g^2C_F}{\pi^3}\int_0^\infty \lambda^4q^3dq  
\nonumber \\
\frac{\tilde M(\lambda^2q^2)}{\lambda^2q^2 + \tilde M^2(\lambda^2q^2)} \left(  \frac{D^0_{p-q}}{\lambda^2}- \frac{D^0_{\sigma-q}}{\lambda^2} \right) .
\eeq

It is easy to find the modified $\tilde M$ that satisfies this rescaled equation. Taking simply $\tilde{M}(\lambda^{2}p^{2})\equiv \lambda M(p^{2})$ we indeed recover Eq.~(\ref{subtractedDSE}) so if $M$ solves the former, $\tilde M$ solves the newer, rescaled one; and
the corresponding relation for the constituent masses is simplest,
\begin{equation} \label{masaescala}
M(0)=\frac{\tilde { M }(0)}{\lambda}\ .
\end{equation}

We put this scaling property of the rainbow DSE to use immediately. Taking $\lambda$ as the ratio of saturation points where 
$\alpha_s=0.4/C_F$,
\begin{equation}
\frac{\sigma_{group}}{\sigma_{SU(3)}}=\lambda,
\end{equation}
the mass function rescales in the same way:
\begin{equation}
\frac{M_{group}(0)}{M_{SU(3)}(0)}=\lambda\ ,
\end{equation}
or simply put, eliminating the auxiliary $\lambda$,
\be \label{rescaleM}
\frac{M_{group}(0)}{M_{SU(3)}(0)} = \frac{\sigma_{group}}{\sigma_{SU(3)}} \ .
\ee
This is a central result. When combined with the exponential growth of the saturation point in Eq.~(\ref{saturationpoint}), we obtain our advertised dependence of the fermion mass with the fundamental dimension of the group under which it is charged, the exponential in Eq.~(\ref{mainresult}) for moderate $N_c$.
This is the reason why fermions charged under a large group are expelled from the low-energy spectrum, all things being equal at the GUT scale.

\begin{figure}
\centerline{\includegraphics*[width=8cm]{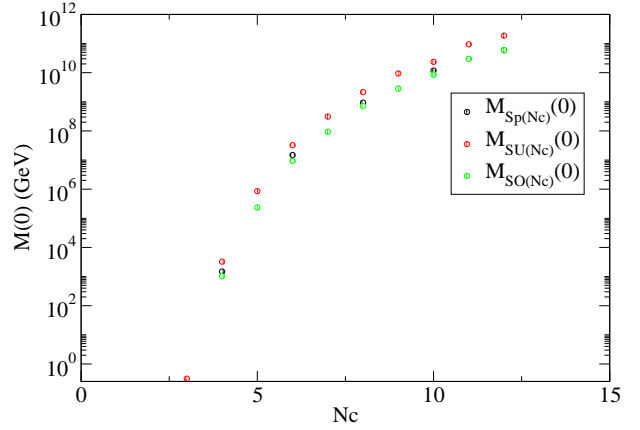}}
\caption{Dynamical mass $M(0)$ as function of $N_c$ from 3 to 12  obtained by 
matching perturbation theory and DSE when 
$(C_F\alpha_s)=0.4$, and obtaining the DSE solution by rescaling that of $SU(3)$.
Because $C_F$ (and more so $C_A$) are basically proportional to $N_c$, there is not much difference between a full non-Abelian theory and a quasi-Abelian truncation in rainbow-ladder approximation, and the scaling is qualitatively similar for the three families of classical Lie groups. 
This stops being true for $N_f$ nearing the critical value, when the exponential Miransky scaling  (see subsection~\ref{subsec:flavor}) of the quasi-Abelian truncation changes to a power-law.) 
 \label{fig:mofNc}}
\end{figure}

Carrying out the rescaling for several values of $N_c$ leads to the 
 dynamical mass $M(0)$ dependence on $N_c$ depicted on figure~\ref{fig:mofNc}.

\section{Discussion and outlook}\label{sec:outlook}

The combination of two methods (perturbation theory and the Dyson-Schwinger equations) has allowed us to show that fermions charged under a large group, if their coupling is equal to the smaller-dimension ones that appear in the Standard Model at the GUT scale $10^{15}$ GeV, are much more massive than the ones we see. In fact, should there exist fermions charged under $SU(4)$ or a group of equal dimension, they would appear in the 10 TeV region, though we cannot pinpoint them to better than order of magnitude estimate because of the crude approximations we have made, but they would not be far out of reach of mid-future experiments.
Perhaps precise calculations in the near future can address this dimension-4 group to predict the mass at which $SU(4)$-charged fermions appear. One can conceive a combination of methods coming together to obtain a good prediction: lattice QCD techniques that have already been demonstrated for groups larger than in the SM~\cite{Bali:2013kia,DeGrand:2015lna,DeGrand:2014cea,Bali:2013fya}, scaling properties of full DSEs or the Exact Renormalization Group Equations~\cite{Bervillier:2014tla,Terao:2000ae,Pawlowski:2005xe}, and multiloop perturbation theory.

It already appears from our simple work that groups yet larger might just endow fermions with a mass not detectable in the foreseeable future.

Should these superheavy fermions be coupled to the Standard Model, they would have long decayed in the early universe due to the enormous phase space available. Were they to exist and be decoupled from the SM, they would just appear to be some form of dark matter.

In addressing the spectrum of Beyond-SM theories one can worry that spontaneous mass generation may break any extant global chiral symmetries and give rise to presumably unseen Goldstone bosons equivalent to QCD's pions.
To dispel doubts, let us recall the Gell-Mann-Oakes-Renner relation~\cite{GellMann:1968rz}
\begin{equation}
M_\pi^2 f_\pi^2 = -2m_q \langle \bar{q}q\rangle
\end{equation}
relating quasi-Goldstone boson mass and decay constant to fermion mass and condensate.
The dependence with the typical scale of symmetry breaking is $f_\pi\sim \Lambda$, $\langle \bar{q}q\rangle\sim - \Lambda^3$, and therefore $M_\pi\sim \sqrt{\Lambda m_q(\Lambda)}$ (note that $m_q(\mu_{GUT})=O({\rm MeV})$, and it is bigger at $\Lambda$). This puts the pseudo-Goldstone bosons out of reach of contemporary experiments, except perhaps for the group $SU(4)$ and equal-dimension ones.
In detailed modeling one can also try to arrange for quantum anomalies lifting the necessity of unwanted Goldstone bosons, such as QCD's $\eta'$. We abstain from attempting this at the present time.

We have shown that the fermion mass for groups slightly larger than $SU(3)$ grows exponentially with $N_c$, because the mass satisfies the same scaling relation than the saturation point, $\sigma$, of the coupling constant
 $ \alpha_{s} $ (which is obviously a proxy for some more sophisticated saturation mechanism), and this point grows exponentially with $N_c$ according to Eq.~(\ref{saturationpoint}).

In our discussion there is a degree of arbitrariness: we have assumed that 
the coupling corresponding to larger groups at the GUT scale, which is totally unknown, is the same for all groups (after all, that is the meaning of GUT). If this hypothesis is lifted, one can of course find arbitrary results. Just like QED with stronger coupling can generate mass spontaneously~\cite{Kizilersu:2014ela}, very large groups with sufficiently small coupling at the GUT scale~\footnote{Due, for example, to sufficiently many flavors screening the interaction.} would not generate it and we would have fermions charged under the oddest groups at current collider scales (which does not seem to be the case).  We also emphasize that our discussion has focused on one or at most few new flavors. If a large $N_c$ is accompanied by a very large $N_f$  one can overcome the gauge-boson antiscreening with fermion screening. Our conclusions then need to be revised.

We insist once more that the couplings for all groups are taken to be \emph{the same} at $\mu_{GUT}$, and we do not suppress them as in t'Hooft's counting~\cite{'tHooft:1973jz} with $g\propto \sqrt{\frac{1}{N_c}}$, which may induce some people to confusion. That counting is a technical device introduced to be able to take the $N_c\to\infty$ limit keeping various quantities, there included the fermion mass, constant (unlike our result); but there is no reason why nature should implement this counting. In fact, the very concept of Grand Unification, hinted at by running coupling constants converging at a high scale, suggests that $g$ would be the same for all groups (that is, independent of $N_c$).

Dynamical mass generation is one of the great conceptual advances of the last half century, turning fermions that are light in the Lagrangian into heavy ones. The phenomenon is well known in QCD and we have discussed groups of larger dimension, through their Casimir factors $C_F$ in the fundamental representation. At the hadron scale we have employed the rainbow approximation of the Dyson-Schwinger equations, and mass generation is approximately proportional to the dimension of the group fundamental representation, with a different slope for each family of classical groups (see figure~\ref{fig:mcutoff}).

In the end, we have provided a plausible answer to the naive question
\emph{Why the symmetry group of the Standard Model,
 $SU(3)_{c} \times SU(2)_{L} \times U(1)_{Y}$, contains only small-dimensional subgroups?}
It happens that, upon equal conditions at a large Grand Unification scale, large-dimensioned groups force dynamical mass generation at higher scales because their coupling runs faster. Since the dynamically generated mass is proportional to the scale at which it is generated, fermions charged under those groups, should they exist, would appear in the spectrum at much higher energies than hitherto explored.

\newpage
\section*{Acknowledgments}
FJLE thanks Richard Williams for useful comments and references at the planning stages of this investigation in 2012, as well as short conversation with Guillermo R\'{\i}os M\'arquez and Feng-Kun Guo.
Work partially supported by the Spanish Excellence Network on Hadronic Physics FIS2014-57026-REDT, and by grants UCM:910309, MINECO:FPA2011-27853-C02-01, MINECO:FPA2014-53375-C2-1-P.

\appendix

\section{Color factors}
\begin{figure}[h]
\centerline{\includegraphics*[width=8cm]{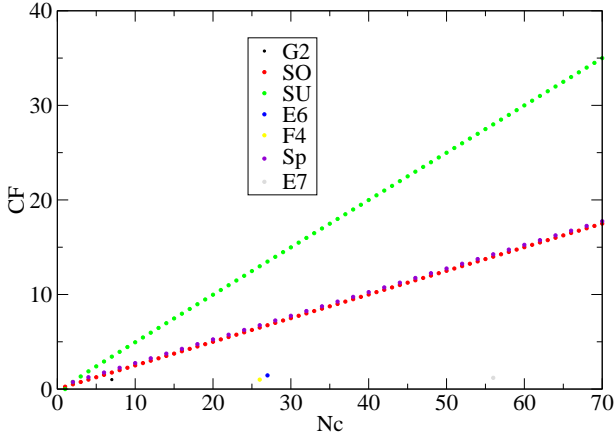}}
\caption{\label{fig:colorfactor} Color factor $C_F$ for the self-energy of a fermion in a gauge theory for the classical groups and the indicated exceptional groups.}
\end{figure}

We need two numbers from group theory, the dimension of the fundamental representation of the group, $N_c$, that is trivially read off, and the color factor $C_F$ for fermion self-interactions, that is calculated in this appendix for various Lie groups. 
There are two classical groups for each odd $N_c$ and three classical ones for each even $N_c$~\cite{Sattinger,vanRitbergen:1998pn}.
Figure~\ref{fig:colorfactor} presents the result at a glance.
It is patent that for each of the classical group families, the relation between $C_F$ and $N_c$ is linear, though with different slope, with the exceptional groups scattered and having a surprisingly small $C_F$ for their large $N_c$.

In an $N_c$-colored  Yang-Mills theory, the $N_{c}$ fermions $q_{i}$ ($i=1,...,N_{c}$) transform in the fundamental representation of an $n$-dimensional Lie group \textit{G}, and the $n$ gauge bosons $ A_{a}$ ($a=1,...,n$), in the adjoint representation.

The $T_{a}$ matrices generate the associated Lie algebra through
\begin{equation}
\left[T_{a},T_{b}\right] =iC_{abc}T_{c},
\end{equation}
with $C_{abc}$ the structure constants. 
To compute each group's color factor in the fundamental representation $C_F$ (its Casimir operator), 
we need to contract $ T_{a}\equiv(T_{a})^{i}_{j} $ and $ T_{b}\equiv(T_{b})^{j}_{k}$ from each vertex~\cite{Peskin}.
Summing over intermediate states $(a,b,j)$,
\begin{equation} \label{contractgens}
C_F\delta^{i}_{k}= \sum_{a,b,j} (T_{a})^{i}_{j} \delta_{ab} (T_{b})^{j}_{k}    =\sum_{a,j} (T_{a})^{i}_{j} (T_{a})^{j}_{k}\ .
\end{equation}

The generators are normalized by
\begin{equation}
Tr(T_{a}T_{b})=\kappa \delta_{ab},
\end{equation}
with $\kappa$ a convention-dependent constant. 
As we wish to generalize the usual $SU(3)$ discussion to other Lie groups,
we fix $T_{a}=\frac{\lambda_{a}}{2}$ with $\lambda_{a}$ the Gell-Mann matrices and then $\kappa=\frac{1}{2}$. 

The result of contracting the generators of Eq.~(\ref{contractgens})
is a sum over a unique set of irreducible tensors for each group, either totally antisymmetric $ f^{ij...k} $, $ f_{ij...k} $ or totally symmetric $ d^{ij...k} $, $ d_{ij...k} $, forming a basis of the corresponding Lie algebra
\footnote{Note that given a tensor $T\in V^{p}\otimes\tilde{V}^{q} $, $V$
being the vector space generated by a basis of  \textit{p} vectors, while 
$ \tilde{V}$ is its dual space generated by the dual basis of  \textit{q} forms; the tensor will have components  $T^{i_{1}...i_{q}}_{j_{1}...j_{p}}$, 
with upper indices denoting covariant, lower ones contravariant components,  and both are related through complex conjugation~\cite{Cvitanovic2}.
}.
We have found the following relations useful for the task,
\begin{equation} \label{fcontraction}
f_{ijm}f^{mjk}= \alpha \delta^{i}_{k}\ ,
\end{equation}
\begin{equation} \label{dcontraction}
d^{ijm}d_{mjk}= \alpha \delta^{i}_{k}\ ,
\end{equation}
with $\alpha$ a normalization constant of the irreducible tensors, due to generalizing those to three or more indices 
(see~\cite{Cvitanovic:1976am}).

Next we will study all classical groups and several exceptional ones, concentrating on the very minimum and most important properties for this calculation and defining the needed irreducible tensors. The outcome is the factor  $C_F$ for each group as function of the fundamental representation dimension $ N_{c} $; the calculation is doable without resource to the explicit values of the generators and structure constants~\cite{Cvitanovic:1976am}.

\subsection{Classical groups}
\subsubsection{$SU(N_{c})$}

The unitary groups are usually denoted  $SU(N_{c})$ and for them,
\begin{equation}\label{paraunitarios}
\frac{1}{\kappa} (T_{a})^{i}_{j} (T_{a})^{l}_{k}=\delta^{i}_{k} \delta^{l}_{j} - \frac{1}{N_{c}} \delta^{i}_{j} \delta^{l}_{k}.
\end{equation}

The fundamental representation of $SU(N_{c})$ 
is the set of  $ \left[ N_{c} \times N_{c}\right]  $ unitary matrices 
with unit determinant acting on an $N_{c}$-dimensional complex space ($N_{c}$ fermions, for our purposes). 
The invariant quantities in this representation are the metric $ \delta^{i}_{j} $ and the Levi-Civita tensor of $N_{c}$ dimensions, $ \varepsilon^{ij...k} $. Tracing over $j$ and $l$ in Eq.~(\ref{paraunitarios}), 
\begin{equation}
C_F \delta^{i}_{k}= \frac{1}{2}\sum_{j} ( \delta^{i}_{k} \delta^{j}_{j} - \frac{1}{N_{c}} \delta^{i}_{j} \delta^{j}_{k} ) = \frac{\delta^{i}_{k}}{2} ( N_{c} - \frac{1}{N_{c}} )\ , 
\end{equation}
and finally we reobtain the well-known result
\begin{equation}
C_F= \frac{1}{2} ( N_{c} - \frac{1}{N_{c}} )\ .
\end{equation}
Now we repeat the calculation for other groups used less often in this area of particle physics.

\subsubsection{$SO(N_{c})$}

For orthogonal groups SO($N_{c}$),
\begin{equation}
\frac{1}{\kappa} (T_{a})^{i}_{j} (T_{a})^{l}_{k}= \frac{1}{2}( \delta^{i}_{k} \delta^{l}_{j} - \delta^{il} \delta_{jk} ).
\end{equation}

The fundamental representation of SO($N_{c}$) is the set of $ \left[ N_{c} \times N_{c}\right]$ orthogonal matrices of unit determinant, acting on a complex vector space which is $N_{c}$-dimensional (for our purposes, $N_{c}$ fermions). 

In this representation, the invariant symmetric tensor is $d^{ij}$ (and its inverse $d_{ij}$).
Diagonalizing $d^{ij}$ and rescaling the $N_{c}$ fermion fields $q_{i} (i=1,...,N_{c})$, we can always find a representation where $d_{ij}=\delta_{ij}$. There is no distinction between upper and lower indices (the fermion and its antiparticle), so that the representation is real.
Tracing again over $j$ and $l$, we find
\begin{equation}
C_F=\frac{1}{4}(N_{c}-1)\ .
\end{equation}

\subsubsection{$Sp(N_{c}$) with even  $N_{c}$.}

For the symplectic groups $Sp(N_{c})$ (that have the sign structure of Hamilton's equations and are thus defined only for even $N_c$), 
\begin{equation}
\frac{1}{\kappa} (T_{a})^{i}_{j} (T_{a})^{l}_{k}= \frac{1}{2}( \delta^{i}_{k} \delta^{l}_{j} - f^{il} f_{jk} )\ .
\end{equation}

The fundamental representation of $Sp(N_{c})$ is the set of matrices of dimension $ \left[ N_{c} \times N_{c}\right]$ with even  $N_{c}$
that leave invariant the antisymmetric tensor $ f^{ij} $  (and its inverse $ f_{ij} $), where 
$$f^{ij}= \left(   \begin{tabular}{cc} 0 & 1 \\ -1 & 0 \end{tabular} \right) $$ 
for $N_{c}$=2, or its multidimensional generalization. Tracing once again over $j$, $l$ and employing the relation 
\begin{equation}
f^{ij}f_{jk}=\delta^{i}_{k},
\end{equation}
we arrive at
\begin{equation}
C_F=\frac{1}{4}(N_{C}+1)\ .
\end{equation}

\subsection{Some exceptional groups}
\subsubsection{$G_2$ ($N_c=7$)}

For the real group $G_{2}$
\begin{equation}
\frac{1}{\kappa} (T_{a})^{i}_{j} (T_{a})^{l}_{k}= \frac{1}{2}( \delta^{i}_{k} \delta^{l}_{j} - \delta^{il} \delta_{jk} ) - \frac{1}{\alpha} f^{i}_{\ jm} f^{ml}_{\ \ \ k}.
\end{equation}

The fundamental representation of 
$G_2$ ($N_c=7$) preserves the symmetric $ \delta_{ij} $ and the totally antisymmetric $ f_{ijk} $ tensors. 
It being a real group, $G_{2}$ requires no distinction between covariant and contravariant indices. Tracing over $j$, $l$  and applying Eq.~(\ref{fcontraction}) for the contraction of the $ f_{ijk} $ we obtain
\begin{equation}
C_F=\frac{1}{4}(N_{c}-3)=1\ .
\end{equation}

\subsubsection{$E_6$ ($N_c=27$)}

Next we examine the exceptional complex group E$_{6}$
\begin{equation}
\frac{1}{\kappa} (T_{a})^{i}_{j} (T_{a})^{l}_{k} = \frac{1}{6} \delta^{i}_{k} \delta^{l}_{j} + \frac{1}{18}\delta^{i}_{j} \delta^{l}_{k} - \frac{5}{3\alpha} d^{ilm} d_{mjk}.
\end{equation}

The fundamental representation of $E_6$ ($N_c=27$), 
leaves invariant the totally symmetric $d^{ijk} $ tensor (and its inverse $ d_{ijk} $). Once more, taking the trace over $j$, $l$ and using now Eq.~(\ref{dcontraction}) to contract the $ d_{ijk} $ tensors, we obtain
\begin{equation}
C_F=\frac{1}{12}(N_{c}-\frac{29}{3})=\frac{13}{9}\ .
\end{equation}

\subsubsection{$F_4$ ($N_c=26$)}

We now proceed to the real $F_{4}$ group, for which
\begin{equation}
\frac{1}{\kappa} (T_{a})^{i}_{j} (T_{a})^{l}_{k} = \frac{1}{9}( \delta^{i}_{k} \delta^{l}_{j} - \delta^{il} \delta_{jk} ) - \frac{7}{9\alpha} ( d^{ilm} d_{mjk}-  d^{i}_{\ km} d^{ml}_{\ \ \ j} )\ .
\end{equation}

The fundamental representation of $F_4$ (with $N_c=26$), preserves the symmetric $ \delta_{ij} $ tensor  and also the totally symmetric $ d^{ijk} $ tensor. Again this is a real group, so covariant and contravariant indices need not be distinguished. Taking the trace over $j$, $l$, the fundamental Casimir falls off in two steps,
\begin{equation}
C_F\delta^{i}_{k}=\sum_{j} ( \frac{1}{18}(\delta^{i}_{k} \delta^{j}_{j} - \delta^{ij} \delta_{jk} ) - \frac{7}{18\alpha} ( d^{ijm} d_{mjk}- d_{ikm} d^{mjj} )\ ,
\end{equation}
\begin{equation}
C_F\delta^{i}_{k}=\frac{1}{18}(N_{c}\delta^{i}_{k}-\delta^{i}_{k}) - \frac{7}{18}\delta^{i}_{k}\ ,
\end{equation}
\begin{equation}
C_F=\frac{1}{18}(N_c-8)=1\ .
\end{equation}
(This computation does require use of one explicit value of the totally symmetric tensor $ d_{ijk} $,  namely that $d^{mjj}=0$ vanishes for a repeated index, which does not follow from symmetry alone).

\subsubsection{$E_7$ ($N_c=56$)}

For the complex group $E_{7}$,
\begin{equation}
\frac{1}{\kappa} (T_{a})^{i}_{j} (T_{a})^{l}_{k}= \frac{1}{24} ( \delta^{i}_{k} \delta^{l}_{j} + f^{il}f_{jk} - \frac{2}{\alpha} d^{ilmn} f_{mj}f_{nk} ).
\end{equation}

The fundamental representation of $E_7$ ($N_c=56$) preserves the totally symmetric tensor  $ d^{ijmn} $ as well as the antisymmetric ones
 $f_{ij}$, y $f^{ij}$. Tracing the closure relation over $j$ and $l$,
\begin{equation}
C_F\delta^{i}_{k}= \sum_{j} \frac{1}{48} ( \delta^{i}_{k} \delta^{j}_{j} + f^{ij}f_{jk} - \frac{2}{\alpha} d^{ijmn} f_{mj}f_{nk} ),
\end{equation}
\begin{equation}
C_F\delta^{i}_{k}= \frac{1}{48} (N_{c}\delta^{i}_{k}+\delta^{i}_{k}),
\end{equation}
\begin{equation}
C_F=\frac{1}{48}(N_{c}+1)=\frac{57}{48}.
\end{equation}
(Here it has been sufficient to note that the contraction
 $ d^{ijmn} f_{mj}f_{nk}=0 $ vanishes as the tensors have opposite symmetry.)
The color factors $C_F$ of all the groups studied in this work are collected in table~\ref{tcolor} for ease of reference.

\begin{table}[H]
\centering
\resizebox*{!}{5cm}{
\begin{tabular}{|c||c|}
\hline
Group & Color Factor ($C_F$) \\ \hline
$SU(N_{c})$ & $\frac{1}{2}\Big(N_{c}-\frac{1}{N_{c}}\Big) \ \ \ \forall N_{c}\in \mathbb{N} $ \\  \hline
$SO(N_{c})$ & $\frac{1}{4}\Big(N_{c}-1\Big) \ \ \ \forall N_{c}\in\mathbb{N}$\\ \hline
$Sp(N_{c})$ & $\frac{1}{4}\Big(N_{c}+1\Big) \  \ \ N_{c}=2n \ \ n\in\mathbb{N}$ \\ \hline
$E6$ & $\frac{1}{12}\Big(N_{c}-\frac{29}{3}\Big) \ \ N_{c}=27$ \\ \hline
$F4$ & $\frac{1}{18}\Big(N_{c}-8\Big) \ \ N_{c}=26$ \\ \hline
$G2$ & $\frac{1}{4}\Big(N_{c}-3\Big) \ \ N_{c}=7$ \\ \hline
$E7$ & $\frac{1}{48}\Big(N_{c}+1\Big) \ \ N_{c}=56$ \\ \hline
\end{tabular}}
\caption{Color factors $C_F$ for fermions in the fundamental representation needed for all the groups studied in this work.
}
\label{tcolor}
\end{table} 

\newpage

\section*{Bibliography}

\end{document}